\documentclass{article}
\usepackage{graphicx,amsmath}
\begin{document}
\centerline{\LARGE\bf Asymmetric transport in the bouncer model:}
\centerline{\LARGE\bf mixed, time dependent, noncompact dynamics.}
\vspace{10pt}
\centerline{\large Carl P. Dettmann}
\centerline{\em School of Mathematics, University of Bristol, Bristol BS8 1TW, United Kingdom}
\vspace{5pt}
\centerline{\large Edson D. Leonel}
\centerline{\em Departamento de Estat\'istica, Matem\'atica Aplicada e
Computa\c{c}˜ao, Universidade Estadual Paulista,}
\centerline{\em CEP 13506-900, Rio Claro, SP, Brazil}

\abstract{We consider time-dependence of dynamical transport, following a recent study of the
stadium billiard in which classical transmission and reflection probabilities were shown to exhibit exponential or algebraic decay depending on the choice of the lead or ``hole''.  The system
considered
here is much more general, having a generic mixed phase space structure, time-dependence
of the dynamics, and Fermi acceleration (trajectories with unbounded velocity).  We propose 
an efficient numerical scheme for this model, observe the asymmetric transport effect, and
discuss observed stretched exponential decays.}

\section{Introduction}
Open dynamical systems, which allow escape through a set in phase space (``the hole(s)''), are
of increasing interest.  They describe practical problems of escape from a system and transport
through a system, including in microlasers~\cite{WM}, acoustics~\cite{MLS,WW},
fluid dynamics~\cite{TSPT}, chemical reaction dynamics~\cite{EWW} and astronomy~\cite{WBW},
as well as experiments involving electrons in semiconductors~\cite{NH} or cold atoms confined by laser
beams~\cite{Saif}.
Open dynamics also provides a useful means to investigate quantum chaos~\cite{N},
including random matrix approaches~\cite{PDLOR} and the fractal Weyl conjecture~\cite{LSZ,S,WM},
as well as Poincar\'e recurrence~\cite{AT}, control~\cite{BP} and
nondestructive measurement of chaotic systems~\cite{BD2}.  

A typical approach in open systems
is to take initial conditions distributed according to some probability measure,
and then consider the probability that the trajectories
remain in the system for some (discrete or continuous)
time without reaching the hole(s).  For transport problems this initial measure is supported on one
of the holes, while for escape problems it is mostly within the system itself.  The time-dependence of
these survival probabilities is related to the system dynamics, particularly at late times.
Generally, for strongly chaotic dynamics, the escape is exponential, while for most other
situations (regular, mixed, intermittent) it follows a power~\cite{AT}.  For the transport problem, the
Poincar\'e recurrence theorem (where it applies) requires that the survival probability decay
to zero, while this need not be the case for the escape problem.
In the strongly chaotic case, exponential escape is described mathematically using conditionally
invariant measures~\cite{DY}; for small holes a local escape rate has been shown to exist, where
the rate of decay is proportional to the hole size, and also depends on short periodic orbits covered by
the hole~\cite{BY,KL}.  Other quantitative approaches have found exact coefficients and expansions
for the escape from open circular~\cite{BD1} and stadium billiards~\cite{DG1,DG2},
and more general strongly chaotic dynamics~\cite{BD2}.

Here we consider the approach of~\cite{DG2}, in which it was found for the stadium billiard that
the transmission and reflection probabilities for trajectories entering the billiard through one of
two holes decayed at long times exponentially or algebraically depending on the choice of
entrance and exit hole.
The stadium billiard consists of two parallel straight sides and two semicircular ends, with the
usual billiard dynamics in which a particle moves in straight lines except for mirror-like reflections
from the boundary.  It is an ergodic and chaotic system~\cite{CM},
described as intermittent as the chaos is
interspersed with long stretches of regular motion almost perpendicular to the straight sides, called
``bouncing ball orbits.''  The intermittency means that the survival probability for the escape problem
(uniform initial conditions within the billiard) decay as a constant divided by the time, where the
constant can be written explicitly in terms of the geometrical parameters~\cite{DG1}.  Thus
the exponential transport properties found in~\cite{DG2} for one hole in a straight segment
and one in a curved segment are unexpected; the mechanism is that orbits from
the hole in the curved  segment are blocked by the other hole from
coming too close to the bouncing ball orbits, and so have properties determined by the
chaos characterising the rest of the phase space.

The stadium billiard is however a very special chaotic system, being completely ergodic, a property
which is easily broken by perturbations~\cite{BG}.  Here we consider a much more
general system, the bouncer model, which in common with generic Hamiltonian systems has
a mixed phase space consisting of elliptic islands and (for certain parameter values) a
large ergodic component with chaotic properties.
It is also a time-dependent system and exhibits Fermi acceleration,
so the phase space is not compact or of finite invariant measure.  Fermi acceleration is a
phenomenon in which a particle gains unbounded energy from a moving infinitely heavy
plate (with usually periodic, specifically sinusoidal motion considered).  The original motivation was
the origin of high energy cosmic rays~\cite{F}.  The particle returns to the plate either due to
another plate which is fixed, the Fermi-Ulam model; a gravitational field, the bouncer model considered
here; or a combination of the two~\cite{LLC,LM}.  These problems are generalisations of billiards
due to the hard collisions, however they are time-dependent and (in the original form) one
dimensional.  The phase of the oscillation plays the role of a second dimension, so that the
collision map is two-dimensional, as for two-dimensional billiards such as the stadium.
Fermi acceleration is observed for the bouncer model, but not in the Fermi-Ulam or combined
models for periodic oscillations of the plate.  Time-dependent two-dimensional billiards have
also been much studied, and can exhibit Fermi acceleration; see~\cite{LB,GT}.

At this point we make some remarks about the chaotic properties of systems with Fermi
acceleration such as the bouncer model.  In the Fermi acceleration regime, recurrence
and ergodicity of a single unbounded component of phase space are likely, but not proven
to our knowledge.  The largest Lyapunov exponent as calculated for typical points in this
component are likely to be positive for the discrete dynamics but zero in the continuous
dynamics, as the chaos-inducing collisions are rare when the particle has high velocity.
Properties such as mixing are not uniquely defined for systems with infinite invariant measure.
We will describe the large component of phase space below as ``strongly chaotic'', but it
should be recalled that this has a more limited meaning for systems such as the bouncer model.

The question of escape and transport in systems with generic mixed phase space consisting of
elliptic islands and chaotic sea(s) has been previously considered by many authors~\cite{AT}.
The general consensus is that time dependence of both escape and transport probabilities is
eventually algebraic (although often apparently exponential at short times), but many different
power laws
have been proposed and observed numerically~\cite{AMK}.  Claims of universal
power law exponents have also been made~\cite{CK}.  We do not observe a universal law, perhaps
due to the small number of prominent islands considered in our system (for fixed velocity
range; see below).  Understanding these exponents, and the question of a universal law
are interesting, important but also very
challenging problems, and we will postpone them to future work.
What we can say about the system considered here is that clear
stretched exponential decay is observed, at least for finite times and in some situations; this
is the subject of discussion below.  The main aim of the paper is elucidation of the transport
properties of the bouncer model, with its characteristics of mixed phase space, time dependence
and unbounded phase space, characteristics of extremely general systems.

In Section~\ref{s:model} we define the bouncer model, and propose a new numerical approach
for efficient solution of its transcendental equations.  In Sec.~\ref{s:esc} we consider the escape
problem and in Sec.~\ref{s:trans} the transport problem.  Concluding discussions are in
Sec.~\ref{s:concl}.

\section{The bouncer model and its numerical simulation}\label{s:model}
In the bouncer model a plate moves periodically with
vertical position $y_0(t)=\epsilon\cos wt$, and a particle with position $y(t)$
makes elastic collisions, returning to the
plate due to a constant gravitational force $-g$.  We can scale the time to set $w=1$ and the
position to set $g=1$, leaving only a single parameter $\epsilon$; here we mostly do not consider
air resistance or an inelastic restitution coefficients, although these have been considered in
the literature~\cite{L,LL,LOC}.  The natural dynamical variables are the
phase $\phi_n=t_n  \pmod{2\pi}$ of the collision at time $t_n$ and the velocity $v_n=\dot{y}(t_n)$ immediately after this collision. The dynamics is given implicitly by
\begin{eqnarray}
\epsilon\cos \phi_{n+1}=\epsilon\cos \phi_n+v_n\Delta_n-\frac{\Delta_n^2}{2}\\
v_{n+1}=-(v_n-\Delta_n)-2\epsilon\sin \phi_{n+1}
\end{eqnarray}
where $\Delta_n=t_{n+1}-t_n=\phi_{n+1}-\phi_n+2\pi k$ with $k\in\{0,1,2,\ldots\}$ is the time difference
between collisions and takes its smallest possible positive value when there is more than
one positive solution.

It is worth mentioning one of the few exact solutions of this model, which also determines the
main phase space structure.  It is $\phi_n=0$,
$v_n=\pi m$, $\delta_n=2\pi m$ for some $m\in\{1,2,3,\ldots\}$, giving a family of fixed
points of the collision map.  Since it corresponds to the top (or for $\epsilon<0$, equivalent to
$\phi_n=\pi$, the bottom) of the cycle,
a linear perturbation of the phase leads only to a quadratic perturbation of the height $y$ and
may be neglected for a linear stability analysis.  It is easy to show that linear perturbations
$(\delta\phi,\delta v)$  satisfy
\begin{equation}\label{e:fp}
\left(\begin{array}{c}\delta\phi_{n+1}\\\delta v_{n+1}\end{array}\right)=
\left(\begin{array}{cc}1&2\\-2\epsilon&1-4\epsilon\end{array}\right)
\left(\begin{array}{c}\delta\phi_n\\\delta v_n\end{array}\right)
\end{equation}
where the matrix has complex eigenvalues corresponding to an elliptic fixed point in the
region $0<\epsilon<1$, in particular only when the particle collides at the top of the cycle
and the forcing is not too strong.   For $\epsilon<0$ or $\epsilon>1$ the orbit is hyperbolic.

Now, let us consider numerical simulation of the bouncer model.
The dynamical equation is transcendental.  The most common approach considered in the
literature to simulate this system rapidly is to simplify it using the Holmes method which uses
a constant
height for the plate, but time-varying impulse provided to the particle; see~\cite{LM}.
Here we consider the exact equations, using the following
efficient numerical method, which is based on the approach used by one of the authors
in  a different billiard problem~\cite{DMR}.  The free flight motion is exactly solvable for a
known time step.    The time step is chosen using a rigorous
lower bound for the actual time step, thus the first collision is never overstepped, modulo
round-off error.  While in this work we consider only the original bouncer model, the
numerical method is discussed in the more general context of an additional frictional force,
which is considered elsewhere in the literature~\cite{L,LL}.

We know $\ddot{y}_0(t)\leq\epsilon$, thus the second derivative of the displacement of
the particle above the plate $d(t)=y(t)-y_0(t)$ satisfies $\ddot{d}(t)\geq -1-\epsilon$
in the absence of friction, where the $1$ corresponds to gravity. 
More generally we consider a frictional force $-m\eta(v)$ given by a monotonically
increasing function $\eta(v)$ with $\eta(0)=0$.  During a free flight the velocity is
decreasing, so the acceleration of the particle is becoming less negative and we have
$\ddot{d}(t)\geq -1-\eta(v(0))-\epsilon$ for $t>0$ within the same
free flight.  We can integrate this inequality twice, using the initial conditions $v(0)$
and $y(0)$ and find in the general case
\begin{equation}
d(t)>d(0)+\dot{d}(0)t-[1+\eta(v(0))+\epsilon]\frac{t^2}{2}
\end{equation}
The RHS is positive at $t=0$, thus a rigorous lower bound for the time step is given by
the solution of the quadratic equation RHS$=0$.  Numerically, the form of the solution is
chosen depending on the sign of the linear term, to minimise subtraction error as usual.
The larger of the two solutions of the quadratic is chosen, corresponding to the collision
in the future.  In the event that roundoff error gives a displacement that is slightly negative,
the use of the larger solution also prevents a spurious collision with the lower side of the plate.

Let us assume we are close to a collision, for which the actual time step required is
$\delta t\ll 1$.  The lower bound computed by the algorithm is as shown above.  An
upper bound for $d(t)$ is the same expression, except that in the acceleration term,
$\eta(v(0))$ is replaced by $\eta(v(t))>0$ and $\epsilon$ is replaced by $-\epsilon$.
Thus the displacement $d_0(t)$ is determined
to within a bounded constant times $t^2$, as is the time step in the generic case when the linear
term in the quadratic equation is not too small.  Thus the algorithm typically converges
quadratically, similar to Newton's method, for example.  In practice, the approach to the
collision to a tolerance close to
machine precision takes around five steps, and is then followed by the collision map
transformation of the velocity.

\begin{figure}
\centerline{\includegraphics[width=500pt]{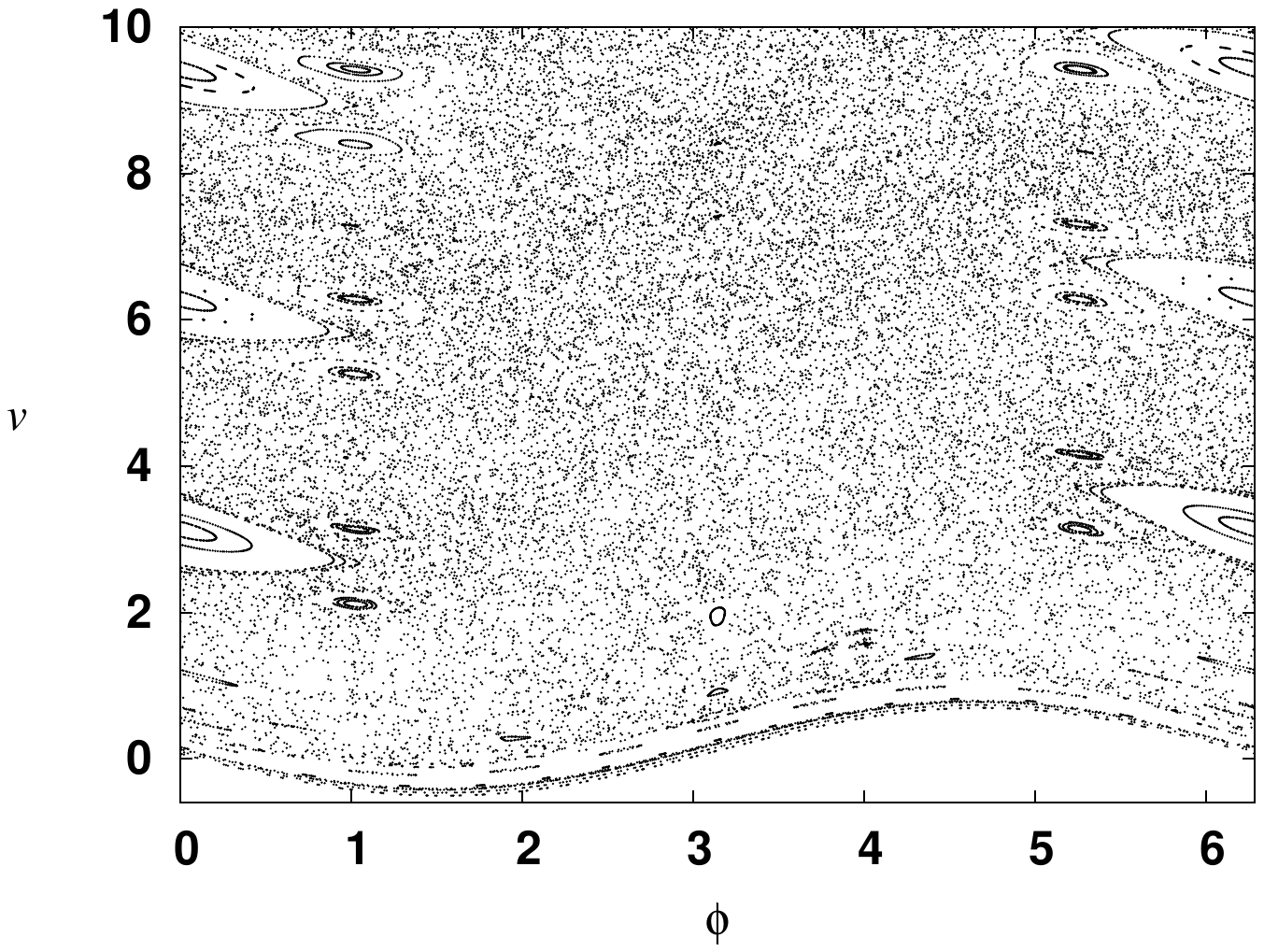}}
\caption{The phase space of the bouncer model for the parameter value $\epsilon=0.6$.
\label{f:phase}}
\end{figure}

The bouncer model exhibits a structure very similar to that of the
Chirikov standard map~\cite{C} with $K\approx 4\epsilon$; see~\cite{LLC,LM}.
This is a 2D map that can be reduced to a torus, being periodic
in both directions.  The bouncer is clearly periodic in $\phi$, and it is also ``nearly''
periodic in $v$.  Recall the family of periodic orbits at $\phi=0$ and $v=\pi m$; clearly both
these and their stability is periodic in $v$ with period $\pi$.  The map as a whole is not
periodic, as can be seen for free flights starting and ending at different $y$ values, but
this becomes less significant at large $v$, and even at small $v$ is strongly noticeable;
see Fig.~\ref{f:phase}.

 The bouncer model exhibits a series of transitions as the parameter is varied, analogous to
 those of the standard map.  The case $\epsilon=0$ is regular. For small $\epsilon$ the
 velocity is bounded due to invariant curves. Near $\epsilon\approx0.24$ (corresponding
 to $K=0.971\ldots$) invariant spanning curves at all but the lowest velocities are disrupted,
 leading to unbounded velocity (Fermi acceleration)
 of some orbits. For $\epsilon>1$ the fixed points studied above become unstable (initially
 bifurcating into elliptic period 2 orbits) and for large $\epsilon$ the motion is almost entirely chaotic.  At some parameter values $K>2\pi$, the standard map also has ``accelerator modes'', which
 are orbits in which velocity increases linearly with discrete time; these can be stable and surrounded by elliptical islands, so yielding anomolous diffusion of velocity for the chaotic component~\cite{IHKM}.
 The parameter $\epsilon=0.6$ of Fig.~\ref{f:phase} shows clear elliptic islands around
 the fixed points, coexisting with Fermi acceleration of the chaotic orbits; this $\epsilon$ is
 below the lowest accelerator modes, so diffusion of velocity in discrete time is expected to be
 normal to a good approximation.
 
 \section{The open bouncer}\label{s:esc}
 Now we discuss the bouncer model as an open system allowing escape, leading to the
 transport problem in the next section.  In open dynamics, we choose a probability measure
 of initial conditions, allow the dynamics to evolve
 with escape, so that any trajectory reaching a subset of phase space called the `hole(s)'
 leaves the system and is no longer considered.  After some time (or number of collisions)
 we ask what fraction of initial conditions are still in the system, the survival probability $P$,
 expressed as a function of $t$ or $n$.
 
 The initial measure can be uniform (which is natural where this is an invariant measure,
 as for bounded Hamiltonian systems), or a more general measure supported in a general
 subset of phase space,  or on the hole.  If the initial measure is supported on the hole, we
 talk about the gap or residence time distribution.  We naturally consider more than one hole,
 and ask for the reflection probabilities (trajectory returns to the same hole) and
 transmission probabilities (escape through a different hole); these are discussed in the next section.
 
 The bouncer model is time-dependent, leading to the question of how to consider open
 time-dependent dynamical systems; general theories of these systems do not appear in
 the literature to our knowledge, although aspects of many specific physical examples have
 been studied.  In the time-dependent case, either the initial probability measure
 or the survival probability function needs to include information about the initial time as well as phase
 space point.  Here we have a periodic system, where the phase can be considered an additional
 phase variable, and it is natural to consider the initial conditions distributed uniformly over
 some interval in  $\phi$ as well as in $v$.  In a quasiperiodically forced system, it would
 be natural to extend the phase space to include a finite number of additional phases.
 For a randomly forced system, the probability measure would naturally incorporate that of
 the forcing.  However for a general deterministic but aperiodic forced system, the best
 approach is likely to be considering the survival probability as a function of both initial
 and final times.
 
 Another issue here, which is common to all non-ergodic systems, and in particular those with
 mixed phase space, is that the results depend on the support of the measure of initial conditions.
 So, if the initial conditions include an elliptic region not connected with the hole, the survival
 probability will tend to a constant as $t\to\infty$, giving the probability of choosing such an initial
 condition.  Numerically, this washes out other time-dependence of $P(t)$ and also slows the
 simulation, as the elliptic trajectories need to be simulated for the full duration of the time interval.
 Ref.~\cite{AT} refers to initial measures that ``touch'' the elliptic island, but this is likely to
 be problematic due to the fractal boundary of these islands.
 Most simulations in the literature therefore use initial conditions in a (largely) strongly
 chaotic region of phase space, and we do this also, drawing $10^8$ uniformly
 from $2<v<4$ and $3.85<\phi<4.15$, and the parameter $\epsilon=0.6$ as in Fig.~\ref{f:phase};
 any other reasonably large set in the chaotic region and any other smooth density would
 give the same results, at least at long times.
 
 The hole is chosen to be a subset of $\phi$ only, ie vertical strips in the phase plot of
 Fig.~\ref{f:phase}.   In general we could
 choose any subset of the phase space.  The choice of intervals in $\phi$ corresponds to
 the plate becoming absorbing at certain parts of its periodic motion.
 In more general terms, this hole has few
 control parameters and typically accesses both regular and chaotic parts of phase space, which
 is the most likely situation experimentally.  This choice of hole location also allows the
 non-escaping orbits to access the Fermi acceleration.  We avoid here the situation where the hole overlaps the set of initial conditions; this belongs to the next section on transport.
 
\begin{figure}
\centerline{
\begin{picture}(400,200)
\put(0,130){\includegraphics[width=200pt]{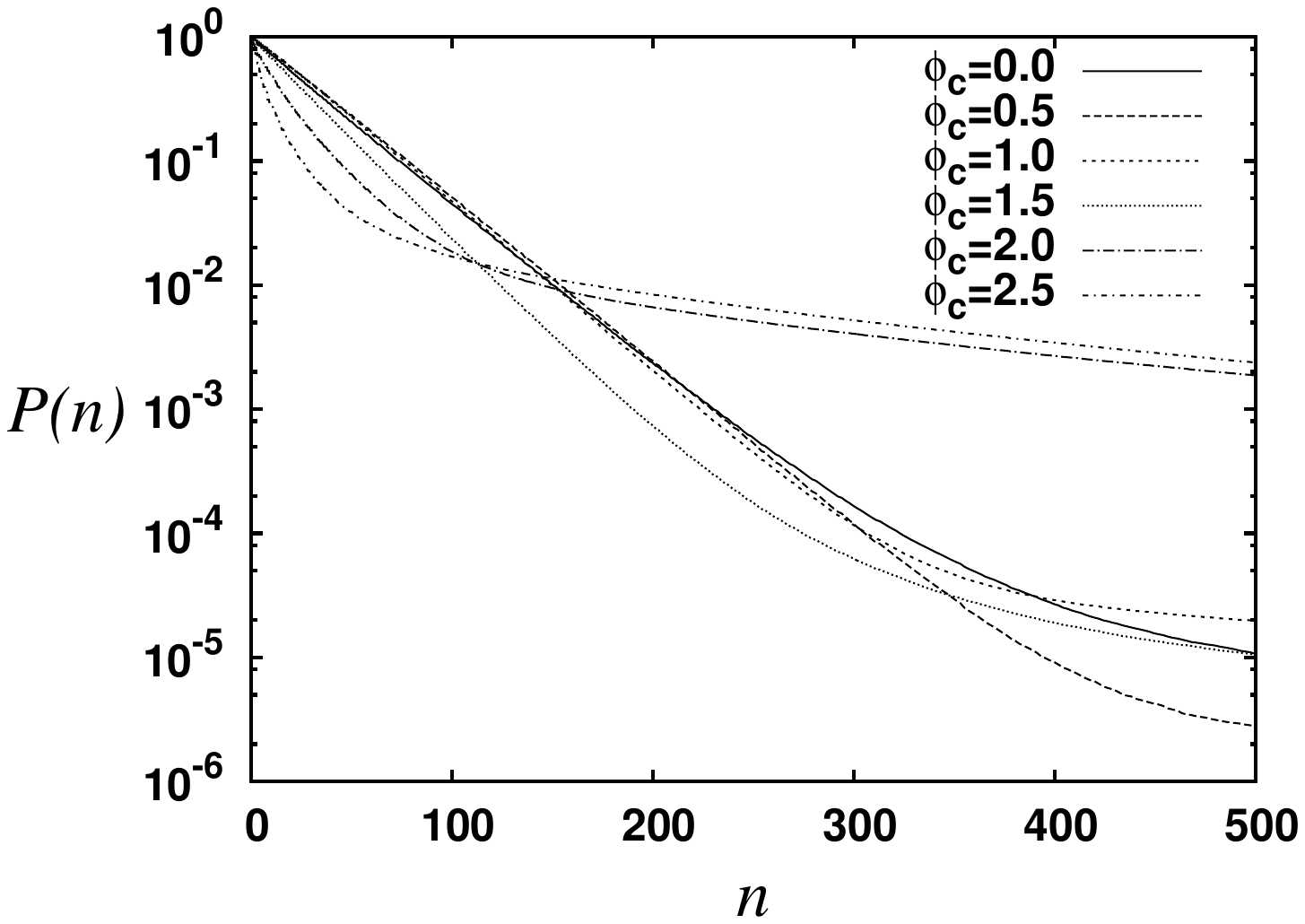}}
\put(173,124){\includegraphics[width=205pt]{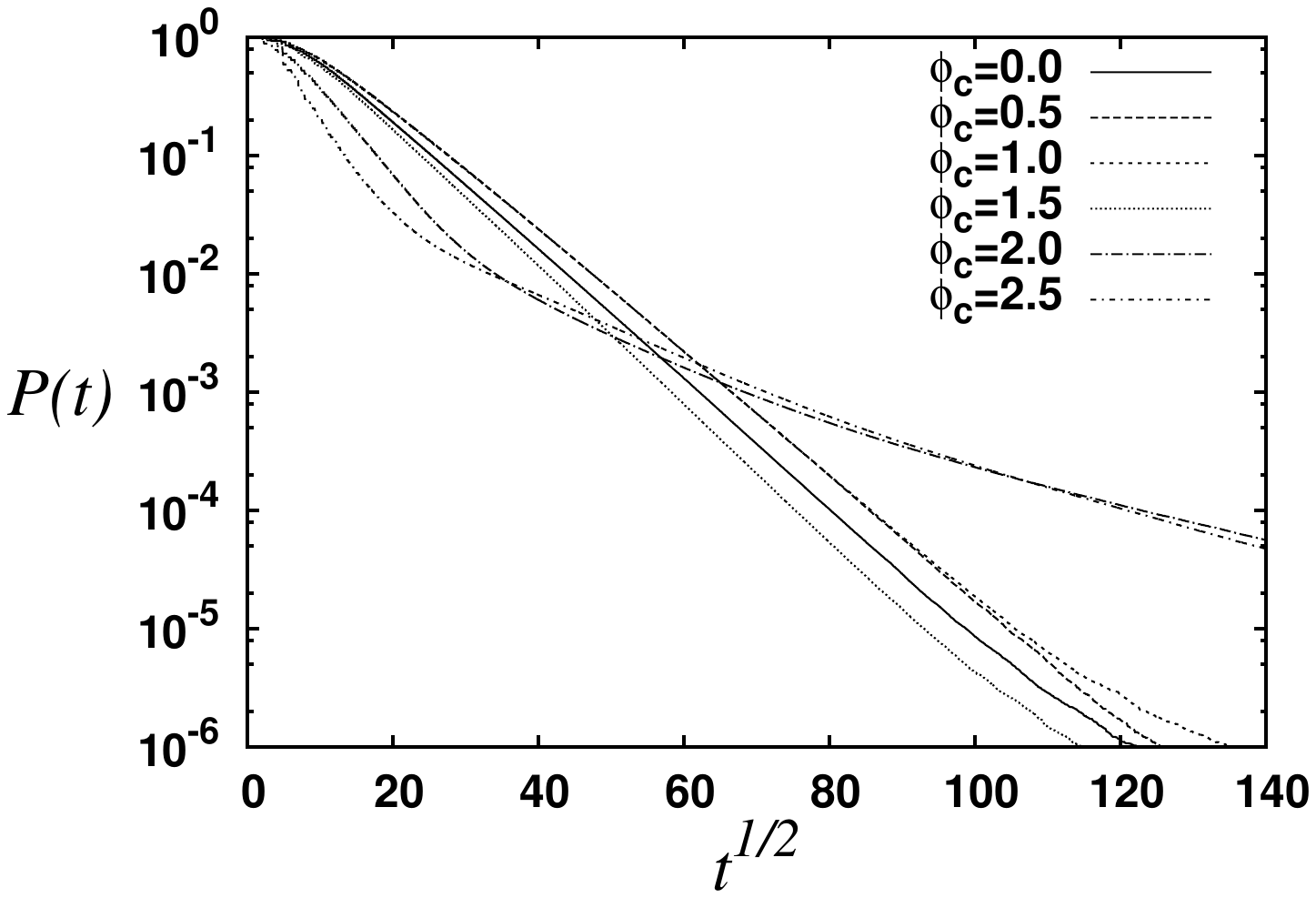}}
\put(0,0){\includegraphics[width=200pt]{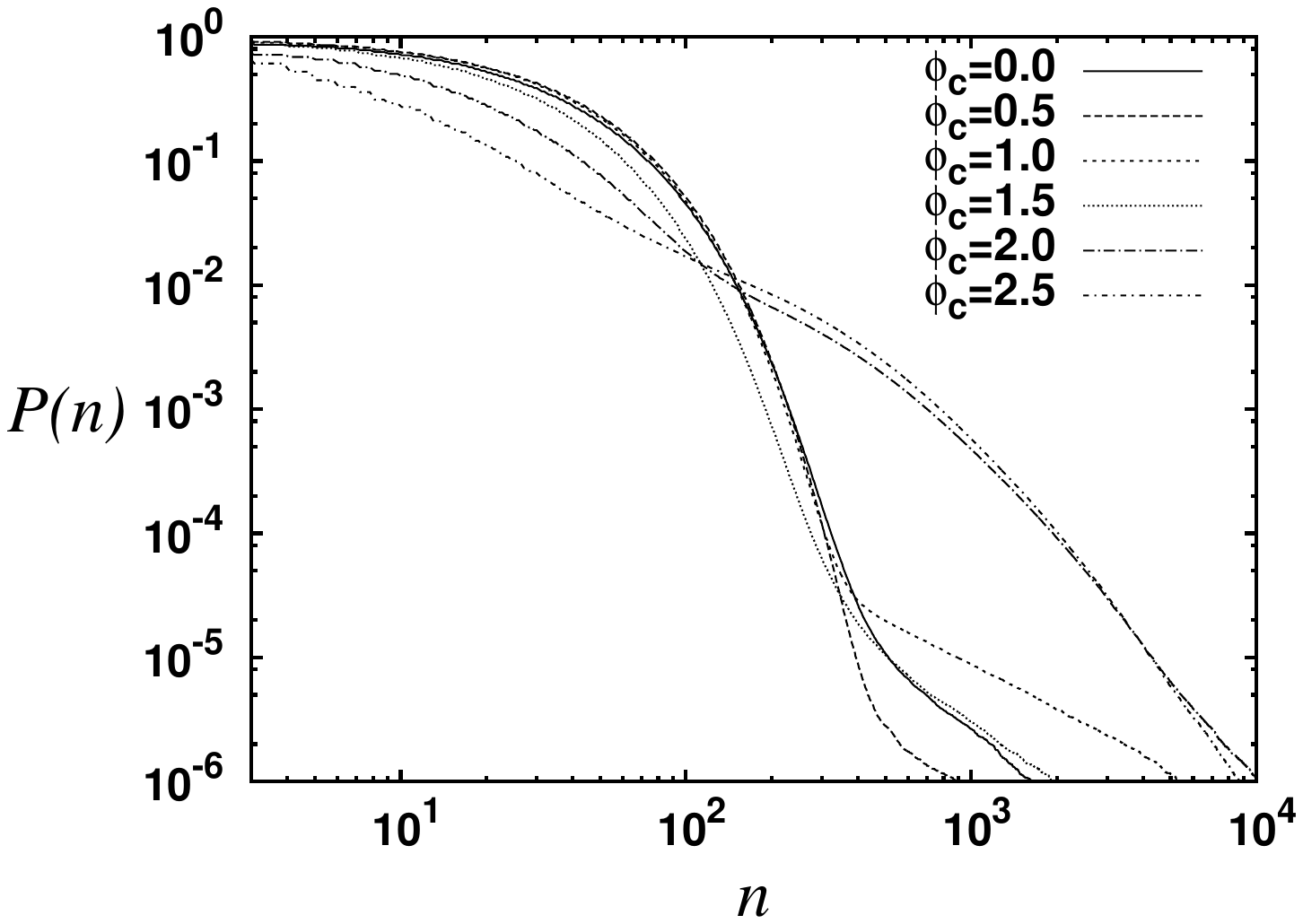}}
\put(177,0){\includegraphics[width=200pt]{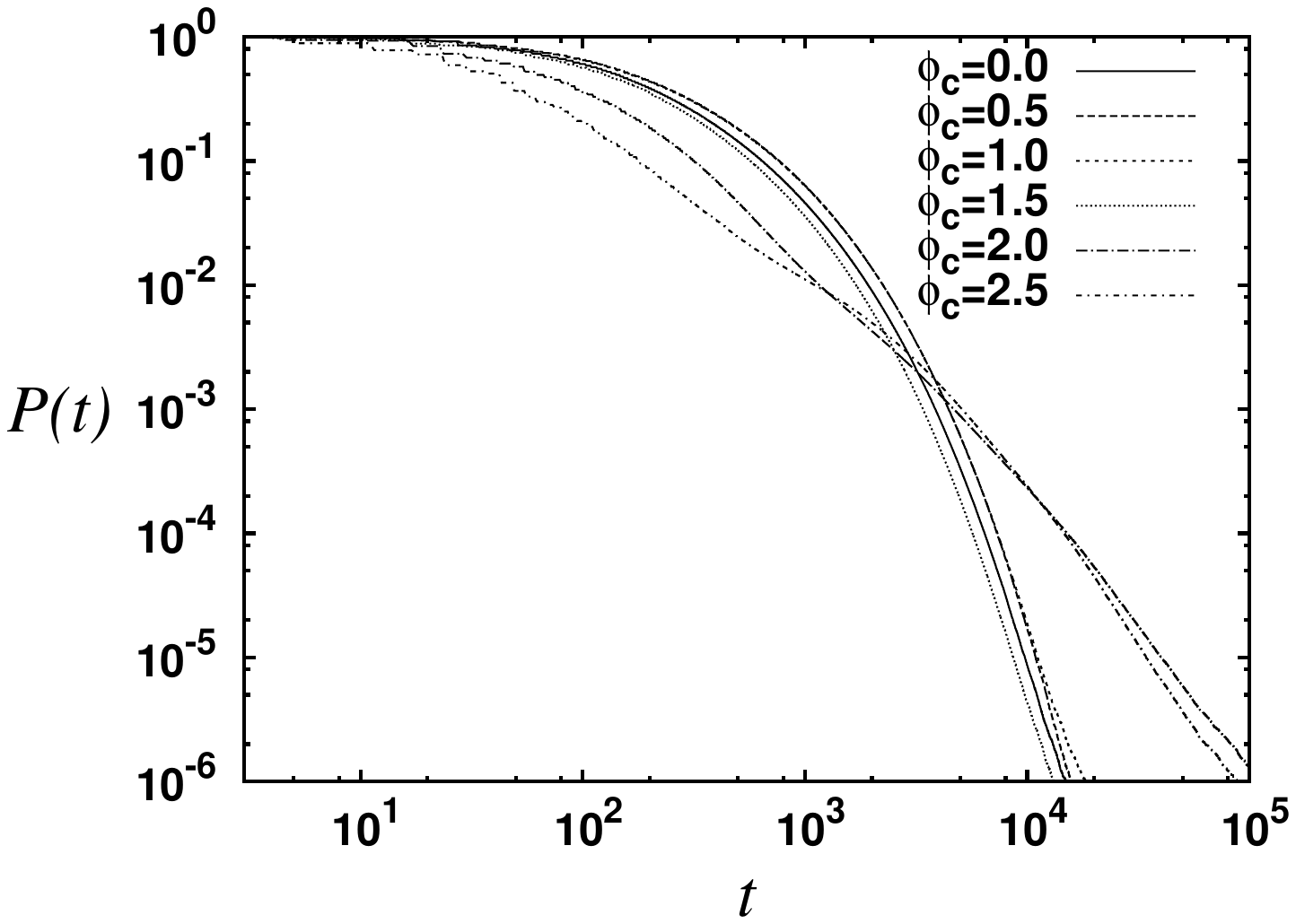}}
\end{picture}}
\caption{Survival probability for the escape problem, for holes of size $h=0.3$ centred at
$\phi=\phi_c$ (compare with Fig.~\protect\ref{f:phase}).  Holes with $\phi_c=0,0.5,1,1.5$
are located over or near the large elliptic islands, while holes with $\phi_c=2,2.5$ are almost
entirely in the chaotic sea.  The two plots on the left are for discrete time, showing primarily
exponential decay in the former case, and algebraic in the latter.
Those on the right are for continuous time, showing stretched exponential decay; note the $t^{1/2}$
on the axis.  The algebraic decay has the same power as for discrete time.
\label{f:esc}}
\end{figure}

 The results for the escape are shown in Fig.~\ref{f:esc}, using a hole of size $h=0.3$ and centred
 at various locations both in and away from the main elliptic islands.  For a hole placed over part
 of the elliptic islands, decay is initially slower, as it takes some time for the dynamics to ``find''
 regions close to the elliptic islands, faster at intermediate times, as the hole covers sticky orbits which would slow decay, and eventually a slow power law, as islands not covered by the hole become
 significant.  Where the hole is in the chaotic sea, decay is generally slowed by the available large
 elliptic islands.  Note that a hole partially covering an island may in fact destroy all orbits surrounding that island; here we see effective destruction of islands using holes quite a lot
 smaller than the islands concerned.
  
 We now focus on the intermediate behaviour in the case where the hole largely blocks orbits
 near the elliptic islands.  When measured in discrete time, this is exponential as expected
 for strongly chaotic systems.  When escape is measured in continuous time (right plots), the
 rapid decay is instead close to a stretched exponential, with $\sqrt{t}$ in the exponent.  This can
 be explained by the Fermi acceleration, as follows: A typical chaotic orbit lasting for $n$ collisions
 has survival probability $e^{-n/n_0}$ where $1/n_0$ is the discrete escape rate.  This orbit diffuses
 in velocity, with a typical velocity proportional to $\sqrt{n}$, and hence time $t_0\approx n\sqrt{n}$ on
 average.  The tail of the distribution of time as a function of $n$ follows a typical diffusion process,
 decaying as $e^{-c_1(t/t_0)^2}$, thus the probability of surviving for $n$ collisions and taking a time
 $t$ is proportional to the product of these exponentials.  For a given time $t$, this is maximised when
 the exponent $-n/n_0-c_1t^2/n^3$ is maximum; differentiating this and setting it equal to zero we find
 the largest contribution from $n\approx(3c_1n_0t^2)^{1/4}$,
 corresponding to a probability $e^{-c_2\sqrt{t}}$ with $c_2$
 determined from the other constants.  The algebraic decays in continuous time are, however, the
 same as in discrete time; this is because it is dominated by orbits near the elliptic island, which do
 not accelerate, and in fact have tightly bounded velocity.
 
 \section{Time-dependent transmission and reflection probabilities}\label{s:trans}
 Recent work~\cite{DG2} has discussed reflection and transmission probabilities in an intermittent
 system, the stadium billiard, as a function of time, and found a striking asymmetry, namely
 that the reflection probability from one hole decayed algebraically, while that from the other
 decayed exponentially, despite the ergodicity of the closed system.  The intermittency in
 the stadium arises from a single family of parabolic ``bouncing ball'' orbits.  The bouncer model
 considered here is
 a significantly more complicated system.  While in the Fermi acceleration regime there is a large
 ergodic component corresponding to the chaotic sea, intermittency arises from the much
 more detailed and generic mechanism of a hierarchy of elliptic (``KAM'') islands.  One of the main
 aims of this paper is to determine whether the same asymmetric effect can occur in this
 much more general setting.
 
 In the stadium of ~\cite{DG2}, the mechanism of asymmetric transport is related to a splitting
 of the ergodic phase space due to the hole in the intermittent region.  We note the hole in
 the straight segment of the stadium also allows escape of orbits with large angles, hence far
 from the bouncing ball orbits, but this does not alter the effect.  The mechanism is that the hole
 ``traps'' the intermittent orbits, so that while they may take some time to escape (due to the
 intermittency), they can escape only through the same hole, and not another hole placed
 far from the intermittent region.
 
 Holes near $\phi=0$ will cover both elliptic
 islands and chaotic sea, while holes near $\phi=\pi$ will cover mostly chaotic sea.  Of course
 any elliptic island completely covered by a hole cannot lead to slow escape.  We must also
 choose a measure of initial conditions.  As we are studying transport, these will be supported
 on one of the holes.  The measure is chosen to be uniform in the $(\phi,v)$ coordinates,
 between $v=2$ and $v=4$, as in the escape case.  Note that the initial measure is supported
 on a strict subset of the hole.  This is because both dynamics and holes have infinite
 measure in this system, so some choice needs to be made about the initial measure; the
 results depend in principle on both the holes (position, size and shape) and on the choice
 of initial measure.  The dependence on the initial measure here is a strength of this
 approach, as it facilitates design of transport problems with specified properties.
 Here our choice is similar to that in the escape section.
 
 For the transport problem we consider $P_{ij}(t)$, the probability of remaining until at least
 time $t$ and eventually escaping through hole $j$ given that the particle enters through
 hole $i$; note the slight difference in notation from~\cite{DG2}.  Here $i,j\in\{1,2\}$ label the
 two holes.  If the particle always eventually escapes (as would follow from Poincar\'e's
 recurrence theorem if the phase space were of finite invariant measure, and is probably
 still true here), we have
 \begin{equation}
 \sum_j P_{ij}(0)=1
 \end{equation}
  
\begin{figure}
 \centerline{
 \begin{picture}(400,200)
\put(0,130){\includegraphics[width=200pt]{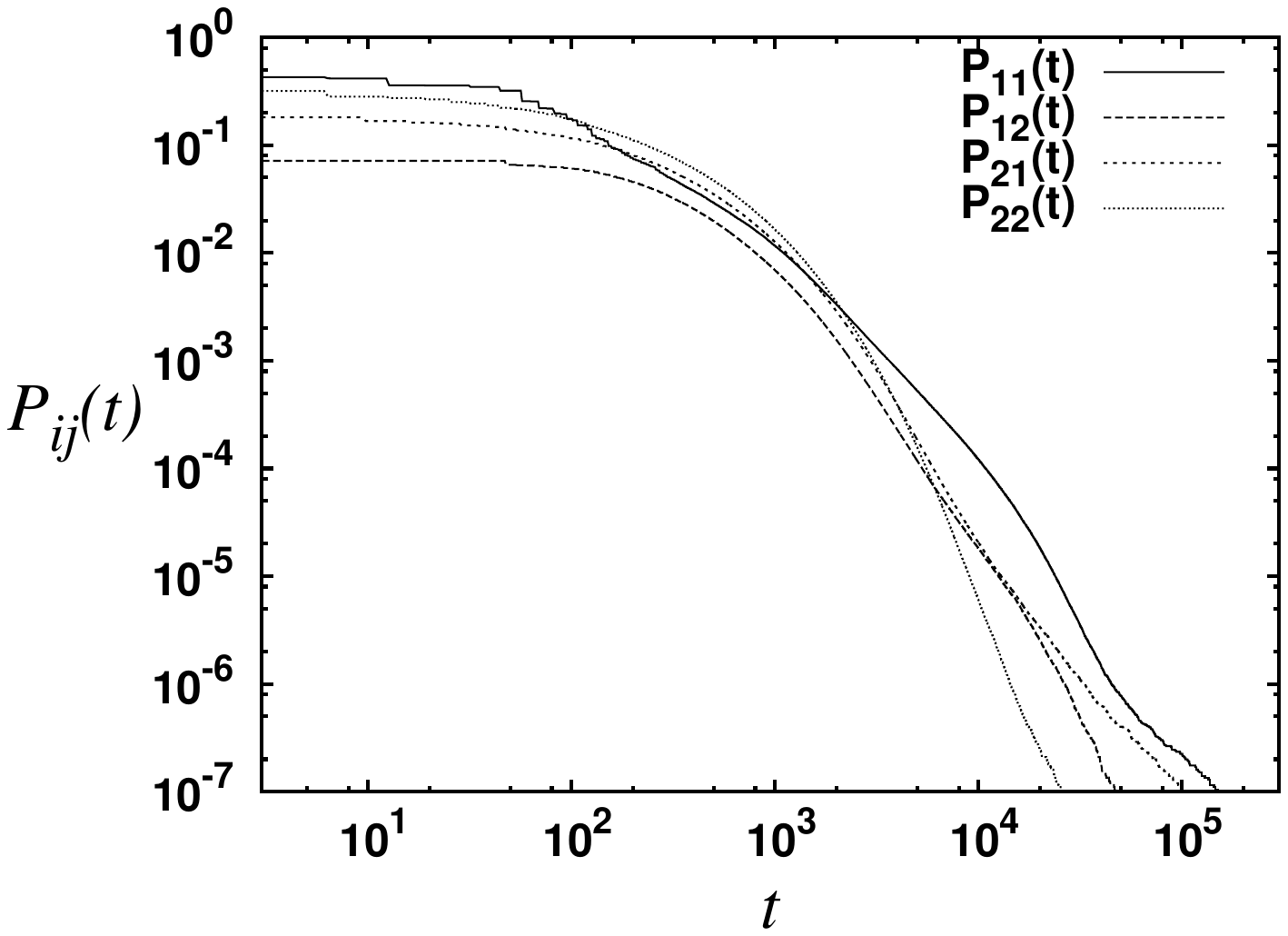}}
\put(175,130){\includegraphics[width=200pt]{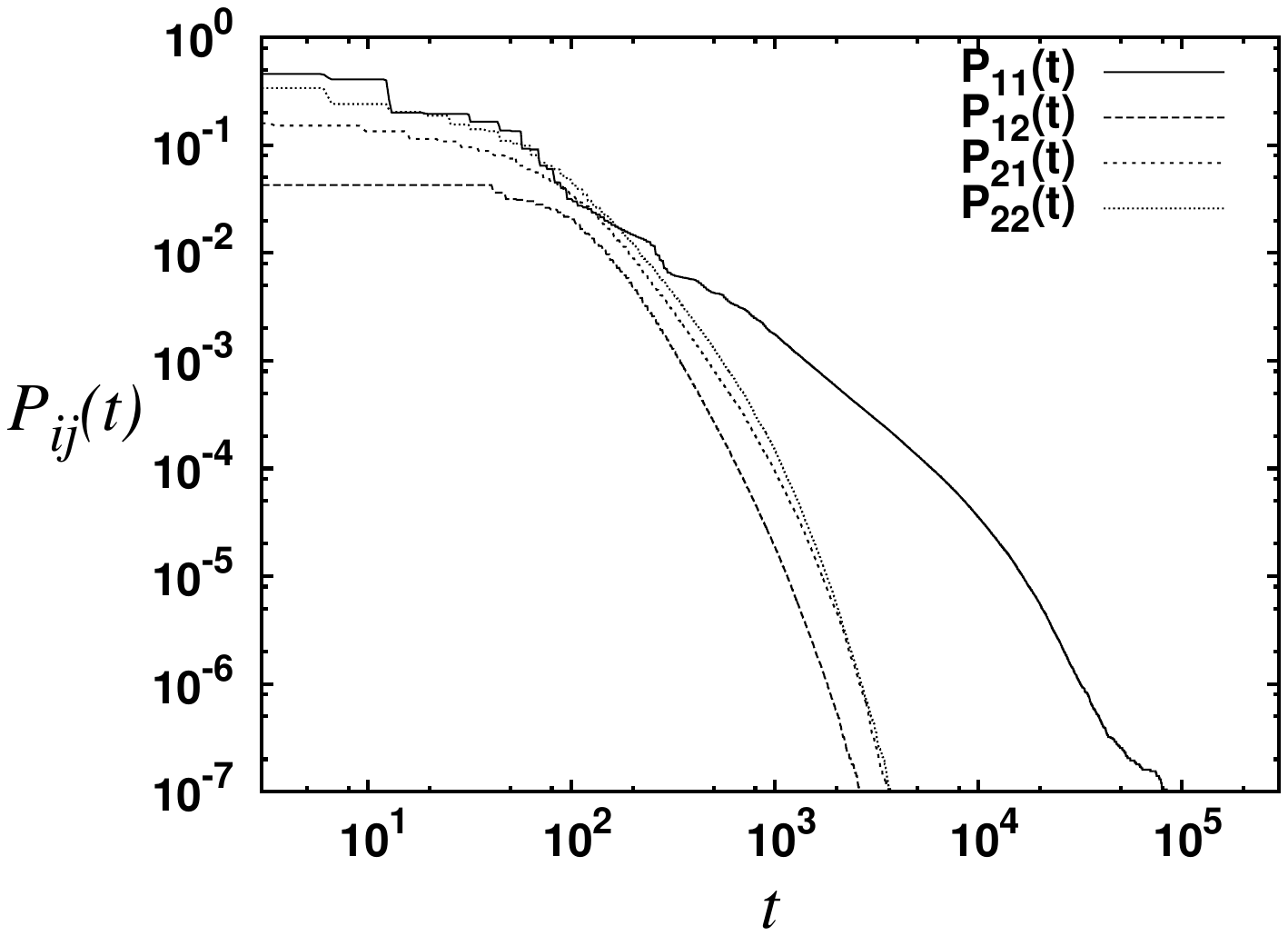}}
\put(0,0){\includegraphics[width=200pt]{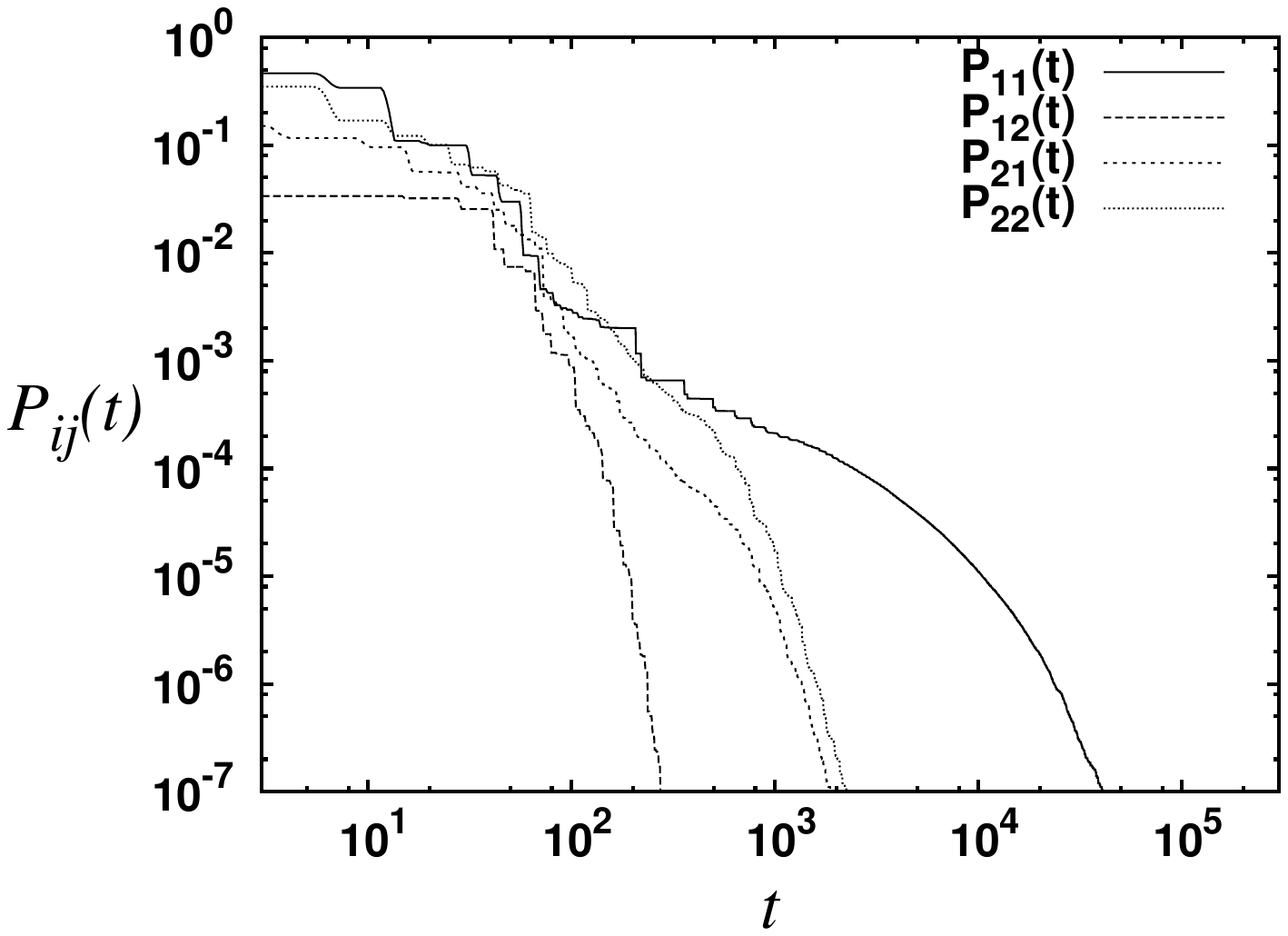}}
\put(175,0){\includegraphics[width=200pt]{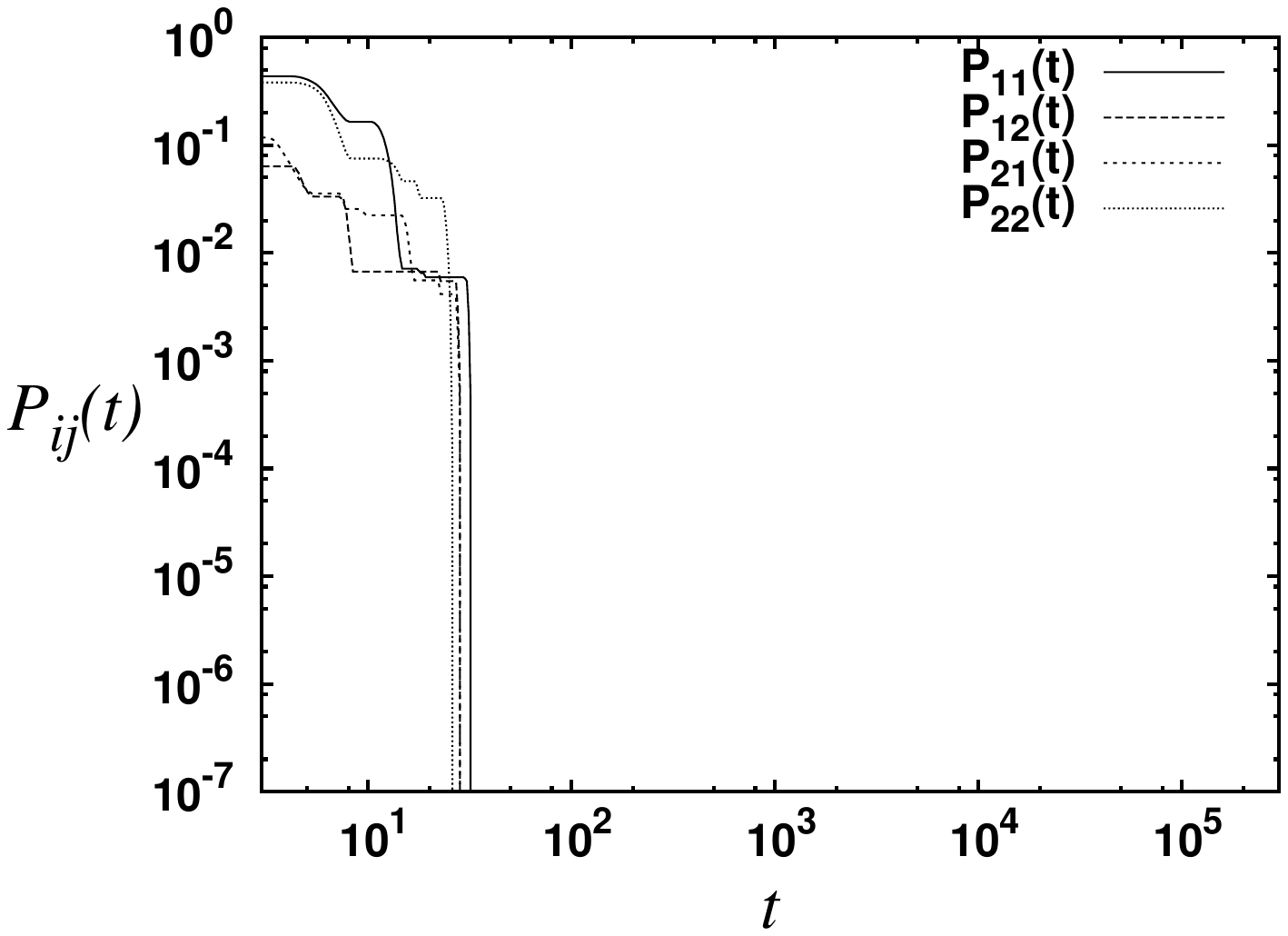}}
\end{picture}
}
 \caption{The transmission and reflection probabilities $P_{ij}(t)$ for four different hole
 sizes $h$: $0.1$ (top left), $0.4$ (top right), $1$ (bottom left) and $2.2$ (bottom right),
 as a function of continuous time $t$. Both holes are the same size.  Hole 1 is centred
 at $\phi=0$ and hole 2 is centred at $\phi=\pi$.
 \label{f:hs}}
 \end{figure}

 For the numerical simulations shown in Fig.~\ref{f:hs}, $10^8$ initial conditions are chosen for
 each hole.  The first hole is in the region of the elliptic islands around $\phi=0$.
 The second hole is centred at $\pi$; this covers some of the smaller islands, and is far from the
 large islands around $\phi=0$, so for practical purposes it is in the strongly chaotic region.
 First we consider the effect of changing the hole size; see Fig.~\ref{f:hs}.  This figure shows
 very clearly the effect we are seeking, namely a very strong $P_{11}(t)$ for hole sizes
 $h=0.4$ and $h=1$.  For these parameter values (and many not shown), the hole in the elliptic
 region effectively traps the long-lived orbits near the smaller islands while reducing the
 lifetime of orbits near the large island.  In other words, orbits near these small islands cannot
 reach other parts of phase space without passing through this hole.  On the other hand, a hole
 that is too small ($h=0.1$) does not effectively trap these orbits, and one that is too large ($h=2.2$)
 allows all orbits to escape within a finite time.
 
 Another feature we can see, particularly in the plot for hole size $1$, is that the long lived
 orbits are well approximated by a stretched exponential, $ae^{b\sqrt{t}}$.  This goes
 against conventional wisdom (see the previous section) in which a power  is expected in
 generic mixed systems, however even in literature including~\cite{AT,EWW}, curves intermediate
 between power and exponential behaviour can be found (but not much remarked upon).  A possible
 explanation is that in situations involving a single large island, the stretched exponential law is
 visible, while for more generic cases with many islands of comparable size, the effects combine to
 generate a more uniform power.  However, in contrast to the escape problem,
 we don't at present have an explanation for why it should be a stretched exponential, or
 whether the power in the exponential is exactly $1/2$.
 
\begin{figure}
 \centerline{
 \begin{picture}(400,150)
\put(0,0){\includegraphics[width=200pt]{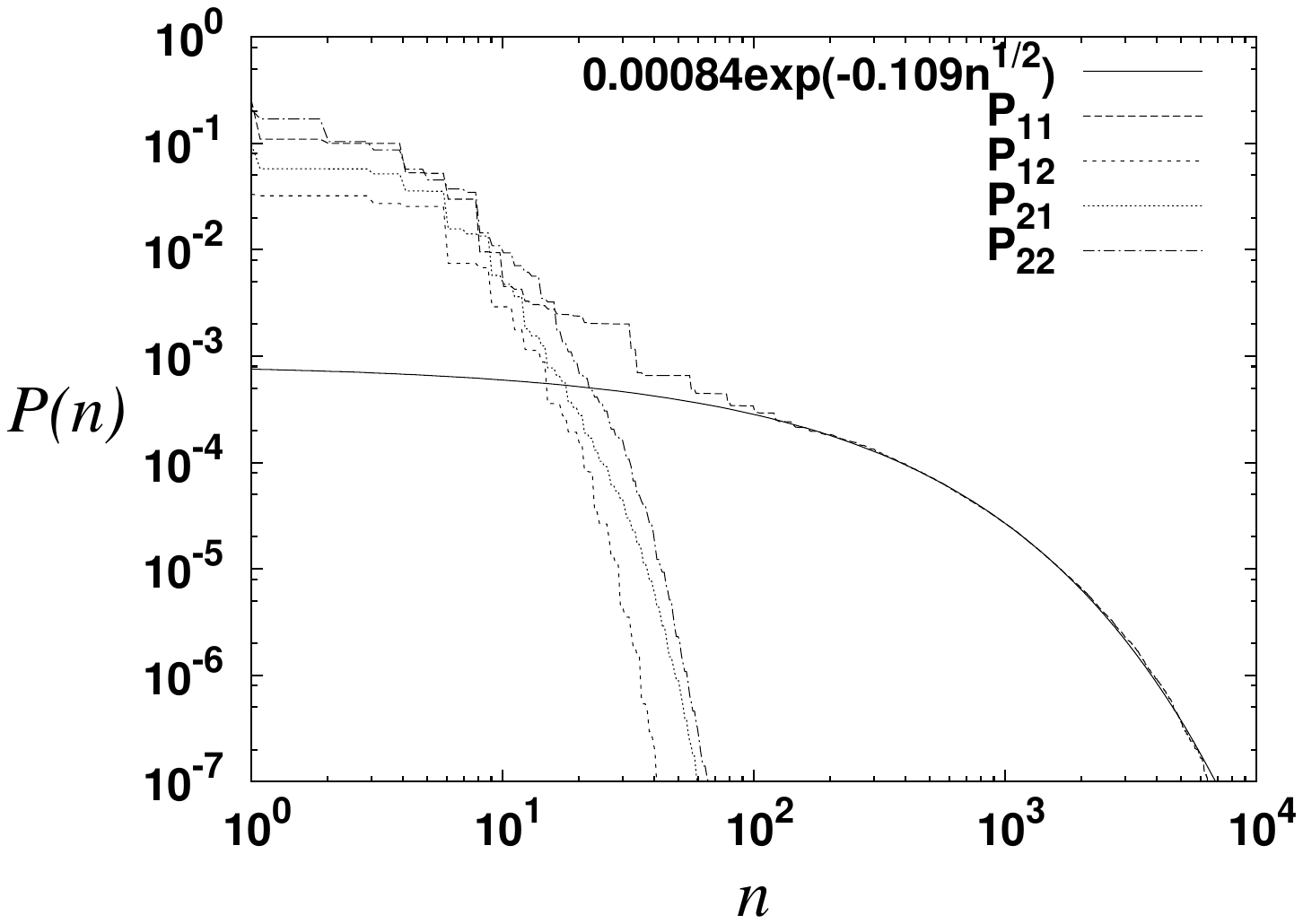}}
\put(175,0){\includegraphics[width=200pt]{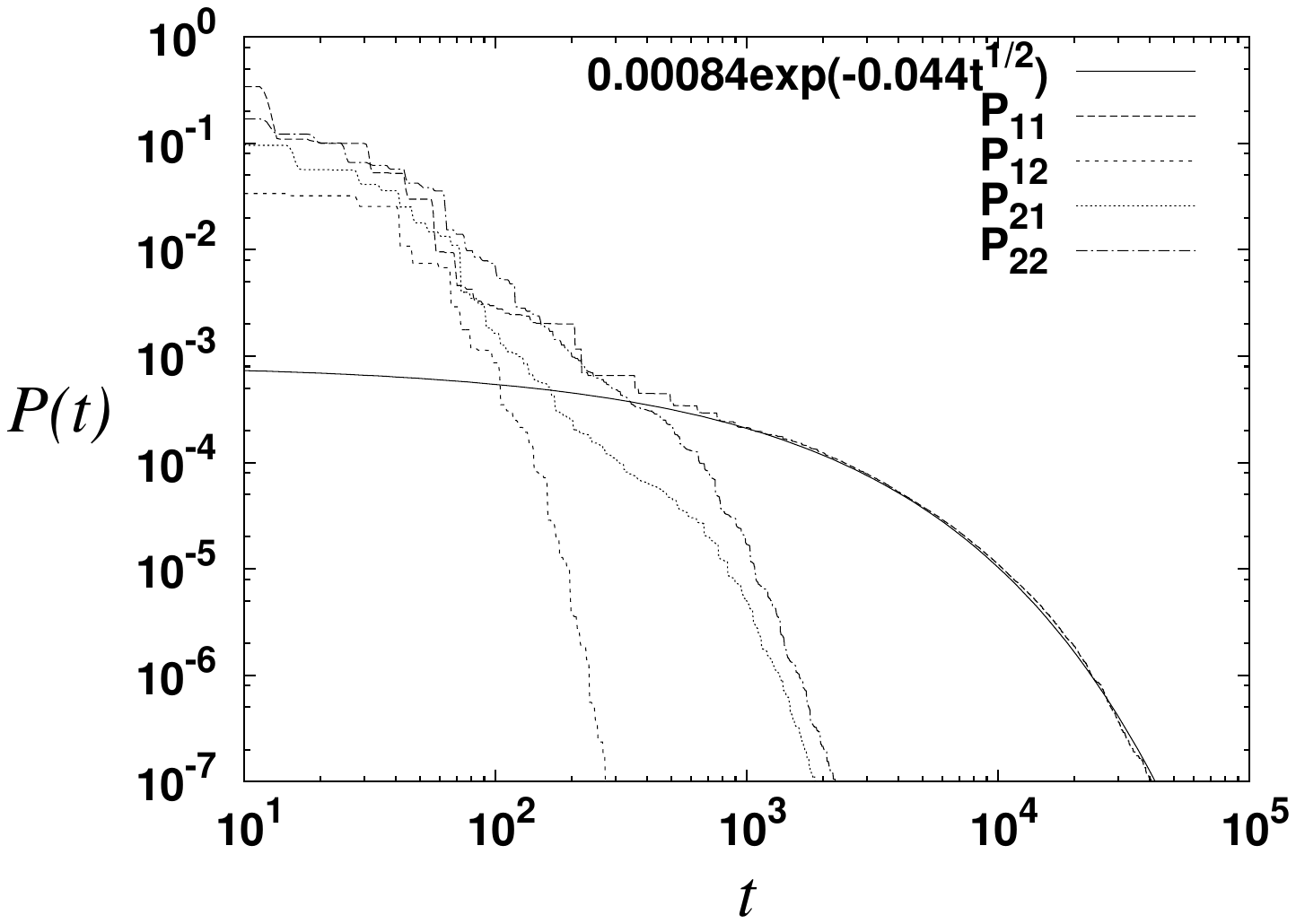}}
\end{picture}
}
 \caption{Parameters as for Fig.~\ref{f:hs}; hole size $h=1$, compared for discrete time
 (left) and continuous time (right), and fits to the stretched exponential behaviour in both cases.
 \label{f:str}}
 \end{figure}

 Fig.~\ref{f:str}, compares transport properties
 of the hole of size $h=1$ in discrete and continuous time, and also shows stretched exponential
 fits for the reflection probability of hole 1 in the region of the elliptic islands.  Here we see
 a difference in the transporting orbits, which decay with many fewer collisions; these can
 reach large velocities.  The discrete time plots for the other values of $h$ are even more similar
 to those of continuous time given in Fig.~\ref{f:hs}, and are not shown.
 
 \section{Conclusion}\label{s:concl}
 In contrast to the stadium billiard considered in~\cite{DG2}, the bouncer model exhibits a generic
 mixed phase space structure, time-dependent dynamics and Fermi acceleration.  The properties
 of escape and transport in mixed phase space have been considered previously, and remain
 a challenging problem for the future.  Here we note that generic mixed phase space allows the
 asymmetric transport similar to that observed in the stadium, but with a different functional form,
 specifically a stretched exponential for the slowly escaping dynamics.
 
 The fact that the dynamics is time-dependent affected the definition of the initial measure and
 survival probabilities, albeit in a fairly straightforward way, that of increasing the phase space
 dimension by one.  Further approaches for situations containing more involved time-dependence
 were discussed in the section on escape.
 
 The Fermi acceleration provided the most interesting and also challenging problems for
 the study of escape and transport.  For the escape problem, it resulted in a different functional
 form of the survival probability between the discrete and continuous time problems, exponential
 in the first case and stretched exponential in the second; the form of the stretched
 exponential (involving $\sqrt{t}$) could be deduced from the properties of the velocity diffusion
 process.  More fundamentally, a hole of infinite measure does not lend itself to a unique
 and natural measure for the initial conditions; this ambiguity and dependence on initial measure
 can be used to advantage, however, in the flexible design of the problem.  This issue deserves further
 study, both from a numerical point of view, and in rigorous infinite ergodic theory.
 
 One or more of these issues apply to very many types of physical systems, and we hope that
 this work will provide a starting point for future studies, both theoretical and experimental,
 discussed in the introduction.  Many such systems are dissipative, and relevant measures
 on chaotic regions of phase space are typically fractal.  Insight into escape and transport
 scenarios is also important for the corresponding quantum problems.
 
\section*{Acknowledgements}  We thank O. Georgiou for a careful reading of this manuscript.
This work was supported by FAPESP, CNPq and Fundunesp,
Brazilian agencies.

\end{document}